\documentclass[sigconf]{acmart}
\AtBeginDocument{%
  }

\setcopyright{acmlicensed}
\copyrightyear{2026}
\acmYear{2026}
\acmDOI{XXXXXXX.XXXXXXX}
\acmConference[Conference acronym 'XX]{Make sure to enter the correct
  conference title from your rights confirmation email}{June 03--05,
  2018}{Woodstock, NY}
\acmISBN{978-1-4503-XXXX-X/2018/06}

\usepackage{multirow}
\usepackage{enumitem}
\usepackage{makecell}

\begin{document}

\def\modelname{ToolRec}
\def\toolkit{SysToolkit}

\title{\modelname{}: Calibrated Preference Alignment for Query Recommendation in On-Device Assistants}

\author{Zihan Luo}
\email{zihanluo@hust.edu.cn}\authornote{This work was done during the internship at OPPO AI Center.}
\affiliation{%
  \institution{Huazhong University of Science and Technology}
  \city{Wuhan}
  \country{China}
}

\author{Lingkui Chen}
\email{chenlingkui@oppo.com}
\affiliation{%
  \institution{OPPO AI Center}
  \city{Beijing}
  \country{China}
}

\author{Ruike Zhang}
\email{zhangruike@oppo.com}
\affiliation{%
  \institution{OPPO AI Center}
  \city{Beijing}
  \country{China}
}

\author{Hong Huang}\authornote{HH is the corresponding author. ZL and HH are affiliated with the National Engineering Research Center for Big Data Technology and System, Services Computing Technology and System Lab, Cluster and Grid Computing Lab, School of Computer Science and Technology, Huazhong University of Science and Technology.}
\email{honghuang@hust.edu.cn}
\affiliation{%
  \institution{Huazhong University of Science and Technology}
  \city{Wuhan}
  \country{China}
}

\author{Boyang Zhang}
\email{z1048826980@gmail.com}
\affiliation{%
  \institution{Huazhong University of Science and Technology}
  \city{Wuhan}
  \country{China}
}

\author{Ziniu Chen}
\email{chenziniu1@oppo.com}
\affiliation{%
  \institution{OPPO AI Center}
  \city{Beijing}
  \country{China}
}

\author{Lizhong Wang}
\email{wlz@oppo.com}
\affiliation{%
  \institution{OPPO AI Center}
  \city{Shenzhen}
  \country{China}
}

\author{Chao Chen}
\email{cschaochen@cqu.edu.cn}
\affiliation{%
  \institution{Chongqing University}
  \city{Chongqing}
  \country{China}
}

\renewcommand{\shortauthors}{Zihan Luo et al.}

\begin{abstract}
  \textit{Large Language Models} (LLMs) have significantly advanced generative query recommendation. However, while alignment is crucial for tailoring LLMs to human preferences, existing alignment methods primarily focus on standard chatbot scenarios, falling short in on-device intelligent assistants where users predominantly expect the rapid invocation of system-level tools. Moreover, directly aligning LLMs with real-world click logs introduces severe noise due to varying user activity levels and the failure to emphasize execution-oriented queries. To address these challenges, we propose \modelname{}, a calibrated preference alignment framework tailored for on-device query recommendation. To ground query recommendation with executable tools, we first construct \toolkit{}, a comprehensive repository of 708 system tools, paired with a context-aware tool retrieval mechanism to ensure that the extracted tools closely match the user's intent. A dual-level calibration mechanism is then proposed to refine raw click data, effectively mitigating user behavioral noise by calibrating signals based on user activity (user-level), while simultaneously up-weighting click signals on tool-invoking queries (system-level). Guided by these refined preference signals, we then align the model using a sample-level weighted \textit{Kahneman-Tversky Optimization} (KTO). Extensive online A/B tests on our mobile assistant platform OPPO Xiaobu, which has over 150 million monthly active users,
  demonstrate that \modelname{} can significantly improve Click-Through Rate (CTR) and total clicks volume over strong baselines while maintaining high query relevance.
\end{abstract}

\ccsdesc[500]{Information systems~Information retrieval}
\ccsdesc[500]{Computing methodologies~Artificial intelligence}

\keywords{Large Language Models, Query Recommendation}


\maketitle

\section{Introduction}

\begin{figure}[t]
    \centering
    \includegraphics[width=3.2in]{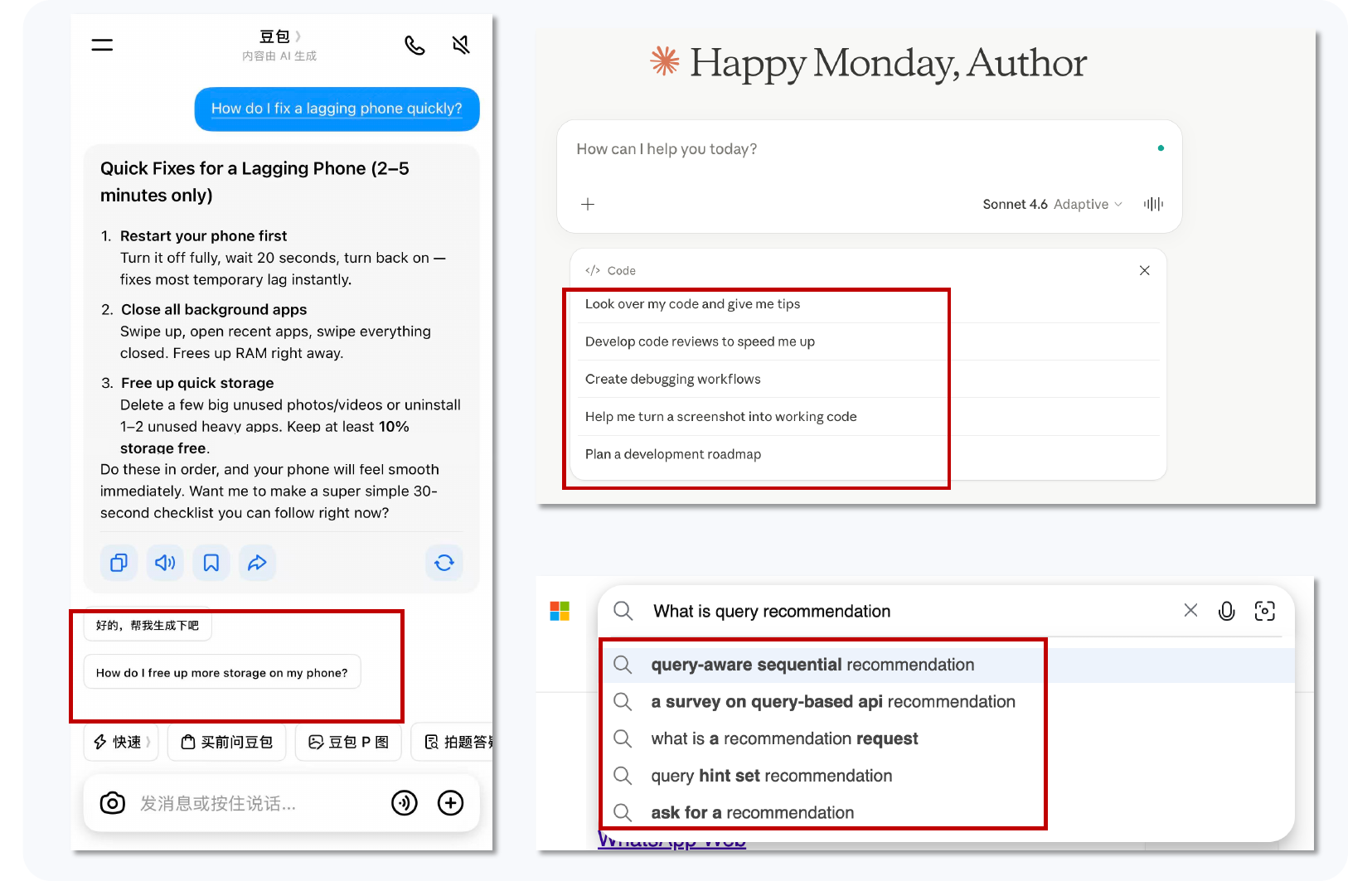}
    \vspace{-0.1cm}
    \caption{Examples of query recommendations in AI assistants and search engines}
    \vspace{-0.3cm}
    \label{fig:interface}
\end{figure}

With the explosive growth of information, humans increasingly rely on search engines and intelligent chatbots to acquire desired information efficiently. In light of this trend, query recommendation has emerged as a vital technique~\cite{linkdin,sp-qac,rrqac}. As shown in Figure~\ref{fig:interface}, by proactively providing users with anticipated query suggestions, it can not only significantly lower the interaction barrier but also effectively improve the user experience and interaction volume.

Recently, driven by the zero-shot generalization capabilities and extensive world knowledge of \textit{Large Language Models} (LLMs), LLM-based generative recommendation has achieved breakthrough progress in the field of query recommendation. For instance, Min et al.~\cite{gqc} utilized \textit{Direct Preference Optimization} (DPO) to align LLM generations with human click behaviors, successfully encouraging the model to output high-quality and diverse query suggestions. Similarly, Yin et al.~\cite{DBLP:journals/corr/abs-2508-15811} introduced Gaussian distributions to capture the uncertainty in user preferences, achieving robust preference alignment via \textit{Group Relative Policy Optimization} (GRPO)~\cite{grpo}.

\begin{figure}[t]
    \centering
    \includegraphics[width=3in]{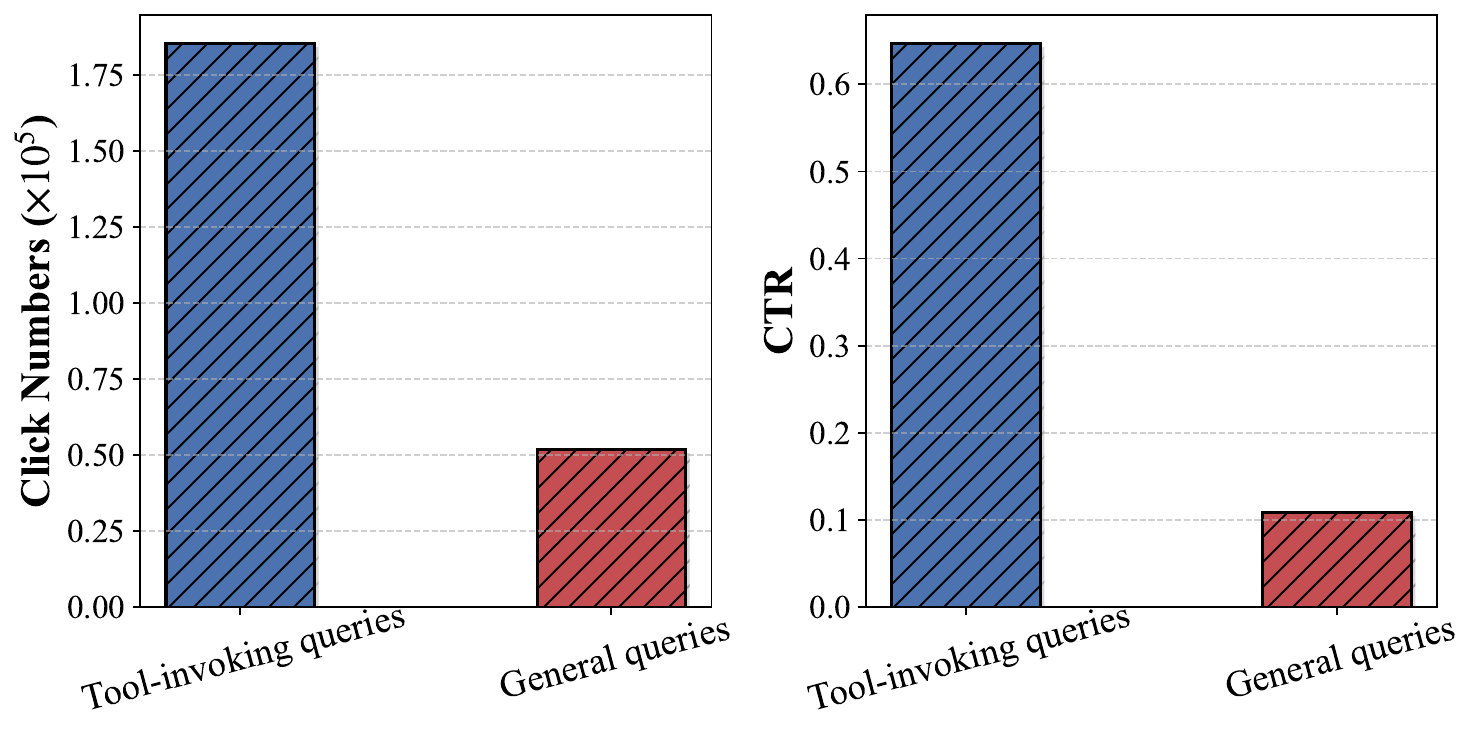}
    \vspace{-0.1cm}
    \caption{The difference in user preferences between tool-invoking queries and general queries in our mobile intelligent assistant OPPO Xiaobu}
    \label{fig:intro}
\end{figure}

Although these methods perform exceptionally well in standard chatbot scenarios, they still fail to capture user inherent preference precisely for on-device intelligent assistants like OPPO Xiaobu. In this scenario, there are two unexplored challenges restricting the practical deployment:
1) \textbf{How to capture the tool-invocation intent in on-device assistants?} Unlike conventional chatbots, a predominant user behavior in on-device assistants is leveraging the assistant to trigger system-level functions on the device. For example, when a user asks "Why is my phone lagging so much?", they typically prefer actionable recommendations like "Clear device cache" that offer immediate system-level utility, rather than passive troubleshooting instructions. Our online data statistics further corroborate this behavioral preference. Specifically, based on whether a recommendation triggers a system-level tool, we categorized all recommended queries into \textit{tool-invoking queries} and \textit{general queries}. As illustrated in Figure~\ref{fig:intro}, statistical results from the past six month demonstrate that both the click-through rate (CTR) and the click numbers for tool-invoking queries are significantly higher than those for general queries. Consequently, delivering high-quality, tool-invoking query recommendations is essential for optimizing user experience and sustaining engagement.
2) \textbf{How to calibrate the varying reliability of implicit preference signals?} Existing alignment methods universally treat click behaviors as golden labels, overlooking the varying reliability of click signals across different scenarios. This introduces significant noise and bias from two perspectives. Firstly, on the user side, there are significant disparities in activity levels among different users. For an extremely inactive user, a "non-click" behavior is largely not due to the poor quality of the generated query, but rather stems from users' weak intention to interact. Thus directly aligning the LLM with such low-quality preference signals may paradoxically lead to suboptimal model performance. Secondly, on the system side, standard alignment fails to distinguish between tool-invoking and general queries. Since a user's authentic preference in on-device scenarios leans heavily towards the rapid invocation of system-level utilities, treating clicks on tool-invoking queries and general queries equally may fail to steer the model toward users' true demands.

To address these challenges, we propose \textbf{\modelname{}}, a novel query recommendation model specifically tailored for on-device intelligent assistants. To equip \modelname{} with the ability to invoke system-level operations, we first construct \textbf{\toolkit{}}, a comprehensive repository containing 708 system tools commonly utilized by users in edge environments. We then introduce a context-aware tool retrieval mechanism, ensuring that only the information of highly relevant tools will be provided to \modelname{} as context, thus maintaining high relevance in tool-invoking query recommendations. Furthermore, to derive reliable and intent-aligned preference data from online logs, we propose a dual-level preference calibration mechanism. Specifically, on the user side, we differentiate the reliability of positive and negative samples based on user activity levels to filter out potential noise. On the system side, we up-weight click behaviors on queries that successfully trigger tools in \toolkit{}—particularly high-frequency ones—to strictly align with users' authentic preference for on-device intelligent assistants. Finally, leveraging the refined preference signals, we utilize a sample-level weighted \textit{Kahneman-Tversky Optimization} (KTO)~\cite{kto} to conduct a more fine-grained preference alignment for the LLM.
The main contributions of this work are summarized as follows:
\begin{itemize}[leftmargin=0.1cm]
    \item We highlight the importance of tool-invoking queries in on-device intelligent assistants and construct \toolkit{}, a comprehensive repository encompassing 708 system-level tools to empower LLM-based query recommendation.
    \item We propose \modelname{}, which incorporates a dual-level preference calibration mechanism from both user and system perspectives, significantly reducing noise in preference data and effectively aligning the model with users' genuine execution-oriented demands via sample-level weighted KTO.
    \item We conduct extensive large-scale online A/B tests on OPPO Xiaobu, a leading mobile assistant platform with over 150 million monthly active users. The results demonstrate that \modelname{} significantly outperforms existing alignment baselines in both CTR and overall click numbers while maintaining high relevance.
\end{itemize}

\section{Related Work}

\subsection{LLMs for Recommendation}
Benefiting from the strong generalization and reasoning capabilities of LLMs, LLM-based generative recommendation is emerging as a new paradigm and has demonstrated remarkable performance. Early works primarily leveraged LLMs for feature enhancement of users and items~\cite{llmesr,flip,llmrec}. For instance, LLMRec~\cite{llmrec} employs LLMs to augment the side information of users and items, providing richer semantic representations for downstream recommendation. FLIP~\cite{flip} aligns multiple modalities at the feature level via LLM-based semantic understanding, and fuses the aligned representations for CTR prediction. Due to the inherent gap between the training tasks between LLMs and recommendation tasks, directly applying LLMs to recommendation often yields suboptimal performance. To bridge this gap, subsequent works align LLMs to user preferences using behavioral signals such as clicks and likes~\cite{sprec,llara}. Early alignment efforts adopt \textit{Supervised Fine-Tuning} (SFT) by formulating next-item prediction as an instruction-following task~\cite{bigrec,tallrec,dlcrec}. More recent work also employs optimization algorithms like DPO~\cite{10.1145/3627673.3679611,sprec} or GRPO~\cite{rankgrpo}, enabling end-to-end generative recommendation models with a stronger alignment to the intent of users. However, all these approaches tend to treat the click behaviors as golden labels, overlooking the varying confidence of click signals across different users.

\subsection{Query Recommendation}

Query recommendation aims to proactively suggest relevant or improved queries based on a user's current input, behavioral signals, or contextual information, thereby facilitating the discovery of more valuable content~\cite{linkdin}. Early approaches to query recommendation primarily took the form of \textit{Query Auto-Completion} (QAC), which retrieves candidate queries from historical search logs by matching a user-typed prefix against logged queries~\cite{sp-qac,rrqac}. A well-known limitation of such methods is their inability to generate suggestions when the input prefix has never been observed in the query log.
To address this issue, subsequent work introduced \textit{Sequence-to-Sequence} (Seq2Seq) models~\cite{nlqac,linkdin}, which endow the system with the ability to generate completions for unseen prefixes and simultaneously enhance the degree of personalization in query suggestions~\cite{linkdin}.
More recently, LLM-based query recommendation methods~\cite{exploratory,gqc,rl4sugg} have emerged as a promising direction, demonstrating strong generalization capability and competitive zero-shot performance. For instance, Liu et al.~\cite{exploratory} leverage in-context learning to prompt LLMs to generate exploratory queries. Besides, RL4Sugg~\cite{rl4sugg}, GaRM~\cite{DBLP:journals/corr/abs-2508-15811} and GQS~\cite{gqc} further incorporate reinforcement learning to align LLMs with user preference signals like clicks.
Distinct from the above works, this paper focuses on query recommendation in the context of on-device intelligent assistants, and further equips the model with tool invocation capabilities to support richer and more actionable interactions.

\section{Preliminary}

\subsection{Problem Formulation}
Given the user's input query $q_u$, the corresponding response from the intelligent AI assistant $\mathcal{A}$, the historical dialogue context $\mathcal{C}$, and the set of available tools $\mathcal{T}$, the LLM-based recommendation model $\mathcal{M}_\theta$ is prompted to generate a set of $K$ candidate queries $\mathcal{Q}_r$ in a single forward pass, which can be formally defined as:
\begin{equation}
    \mathcal{M}_\theta(q_u, \mathcal{A}, \mathcal{C}, \mathcal{T}) \rightarrow Q_r
\end{equation}
where $Q_r=\{q_r^1, q_r^2, \dots, q_r^K\}$, and $K$ is a hyper-parameter predefined by the inference latency limits. 
Before being exposed to the users, the generated candidate queries $Q_r$ will be further processed by downstream reranking and recall modules, which are not the focus of this work, and a set of $N$ candidate queries $Q_r^\prime$ will be exposed to the users at last, where $N$ is determined by the UI slots in advance. 

In practical scenarios, positive and negative samples defined by user click behaviors are widely adopted~\cite{gqc,DBLP:journals/corr/abs-2508-15811}. Let $y(q_r) \in \{0, 1\}$ denote the click indicator for a query $q_r$. The candidate set $Q_r$ is designated as a positive sample, denoted as $Q_r^+$, if and only if $\exists q_r \in Q_r$ such that $y(q_r) = 1$. Conversely, the set is categorized as a negative sample, denoted as $Q_r^-$, if and only if $\forall q_r \in Q_r$, $y(q_r) = 0$.

Our goal is to align the model $\mathcal{M}_\theta$ with the genuine user preferences, thus improving the quality of the recommended queries.

\subsection{Kahneman-Tversky Optimization (KTO)}

Distinct from \textit{Proximal Policy Optimization} (PPO)~\cite{rlhf} or \textit{Direct Preference Optimization} (DPO)~\cite{dpo}-which necessitate paired preference data—the primary advantage of KTO lies in its ability to operate effectively without such paired comparisons. Given a reference policy $\pi_{\text{ref}}$ and a dataset $D$ of prompt-response pairs $(x, y)$, the KTO objective for $\pi_{\theta}$ can be formulated as:
\begin{equation}
    \mathcal{L}_{\text{KTO}}(\pi_\theta, \pi_{\text{ref}}) = \mathbb{E}_{x,y \sim D} [w(1 - v(x, y; \beta))]
    \label{eq:kto_loss}
\end{equation}
where $v$ models a human value function based on prospect theory~\cite{tversky1992advances}:
\begin{equation}
    v(x, y; \beta) =
    \begin{cases}
    \sigma(r(x, y) - z_{\text{ref}}) & \text{if } y \sim y^{+} \\
    \sigma(z_{\text{ref}} - r(x, y)) & \text{if } y \sim y^{-}
    \end{cases}
\end{equation}
where $\sigma$ is an activation function. The implicit reward $r(x, y) = \beta \log \frac{\pi_\theta(y|x)}{\pi_{\text{ref}}(y|x)}$ measures the scaled deviation from the reference model, where $\beta$ represents the KL-divergence penalty. The reference point $z_{\text{ref}}$ can be defined as the expected KL divergence across the dataset, serving as a dynamic baseline:
\begin{equation}
    z_{\text{ref}} = \mathbb{E}_{x' \sim D} [\beta \mathbf{KL}(\pi_{\theta}(y'|x')\|\pi_{\text{ref}}(y'|x'))]
\end{equation}

Finally, the weight $w$ in Eq.(\ref{eq:kto_loss}) applies asymmetric penalties ($\lambda_D$ and $\lambda_U$) for desirable and undesirable outcomes, modeling human loss aversion in prospect theory:
\begin{equation}
    w =
    \begin{cases}
    \lambda_D & \text{if } y \sim y^{+} \\
    \lambda_U & \text{if } y \sim y^{-}
    \end{cases}
\end{equation}

In the context of query recommendation tasks, $x$ represents the user's comprehensive interaction context (including the current input query $q_u$, dialogue history $\mathcal{C}$, and assistant responses $\mathcal{A}$). Meanwhile, $y^{+}$ and $y^{-}$ denote the clicked recommended queries $Q_r^+$ and non-clicked recommended queries $Q_r^-$, respectively.

\section{Methodology}
In this section, we will introduce \modelname{} in detail, a query recommendation model tailored for on-device intelligent assistants.

\begin{figure*}[t]
    \centering
    \includegraphics[width=6.1in]{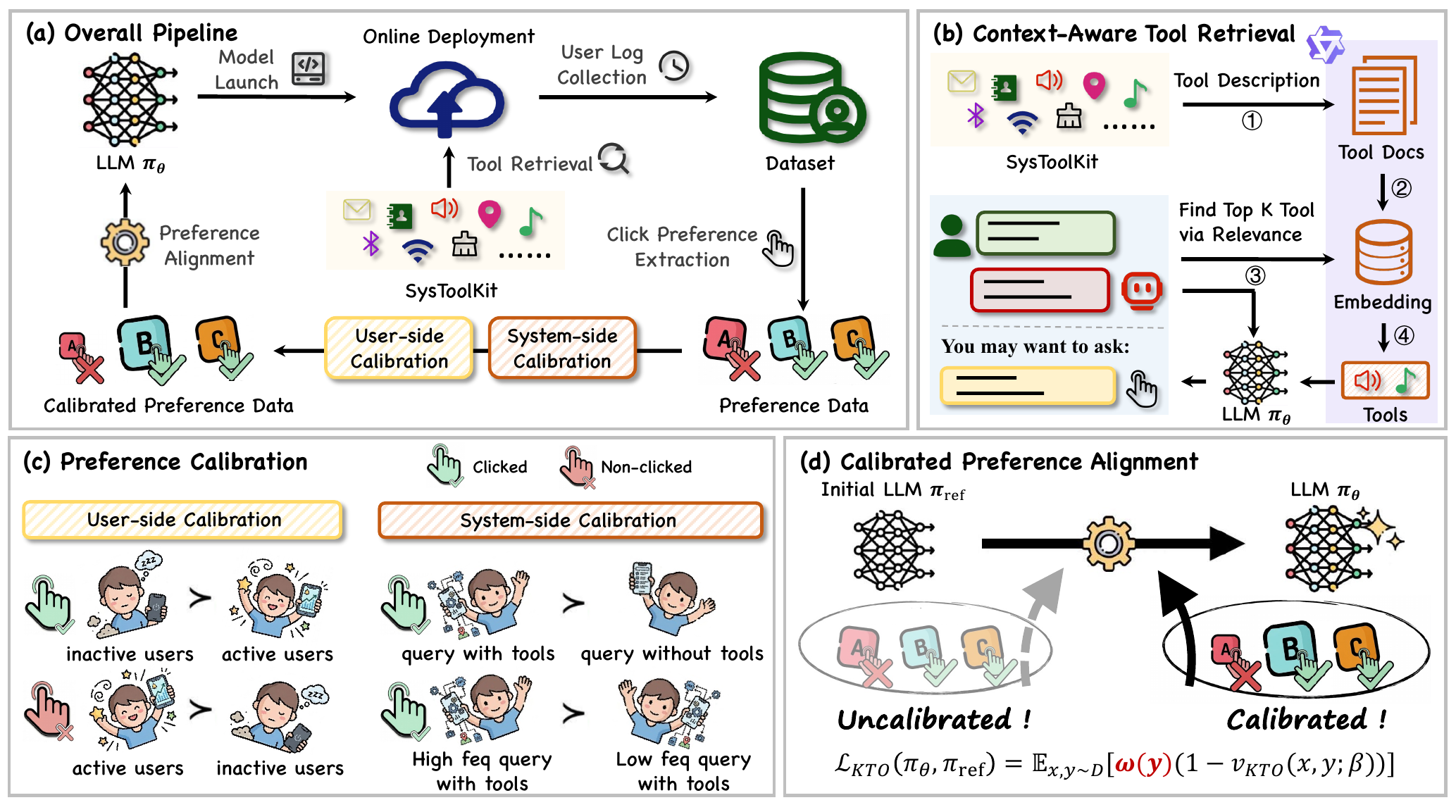}
    \caption{An overview of our proposed \modelname{}. (a) The overall pipeline of \modelname{} establishes a closed-loop system from online deployment to model optimization. (b) During online serving, a context-aware tool retrieval module actively fetches relevant tools from the SysToolKit before generating query recommendations for users. By collecting users' responses (clicks or non-clicks) to these queries, we extract raw implicit preference data. (c) To further improve the preference data quality, these raw click preference data are then refined by a dual-level calibration mechanism, incorporating both user-side calibration and system-side calibration. (d) The resulting calibrated preference data ultimately drives the final preference alignment phase, optimizing the LLM via weighted KTO.}
    \label{fig:framework}
\end{figure*}

\subsection{Overall Framework}
The overall framework of \modelname{} is illustrated in Figure~\ref{fig:framework}. To equip \modelname{} with the ability to invoke system-level functions on the device, we first construct \toolkit{}, a comprehensive repository that includes 708 system tools commonly utilized by users in edge environments. To ensure that \modelname{} maintains relevance when recommending tool-related queries, a context-aware tool retrieval mechanism is then proposed. Specifically, only the information of the top-$K$ highly relevant tools from \toolkit{} is provided to \modelname{} as context. After that, we gather the basic preference data from online logs and click behaviors, and propose a dual-level preference calibration mechanism to further improve the quality of preference data. Specifically, on the user side, we differentiate the reliability of positive and negative samples based on user activity levels to extract preference signals of higher-confidence. On the system side, we up-weight queries that successfully trigger tools in \toolkit{}—particularly high-frequency ones—to strictly align with users' authentic preference for on-device smart assistants. Finally, recognizing the impracticality of acquiring paired preference data for identical inputs in online scenarios, we conduct preference alignment via a sample-level weighted KTO utilizing the calibrated data. The details of each module in \modelname{} will be elaborated in subsequent parts.

\subsection{\toolkit{}: System-level Tool Repository}

Distinct from query recommendation in traditional chatbots or search engines~\cite{gqc}, a core user requirement for on-device intelligent assistants is the rapid invocation of relevant system functions. To enhance the user experience and improve the quality of query recommendations, we introduce \textbf{\toolkit{}}, an encapsulated toolkit designed for \modelname{}. In this part, we will detail the specific design of \toolkit{} from the following two perspectives:

\begin{figure}[t]
    \centering
    \includegraphics[width=2.8in]{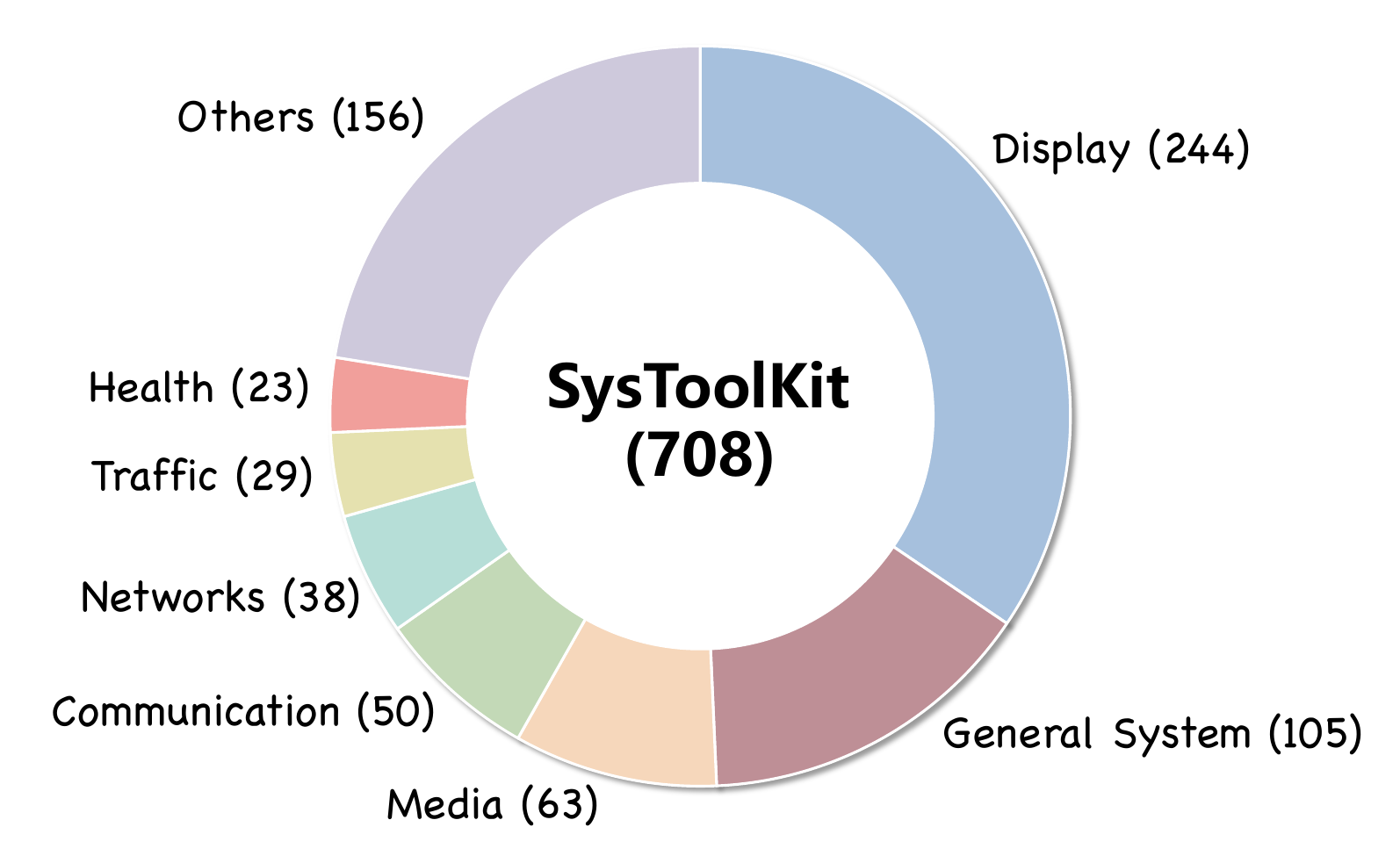}
    \caption{Tool distribution of \toolkit{}}
    \label{fig:toolkit}
\end{figure}

\textbf{Tool Preparation.} 
As mentioned above, the primary objective of \toolkit{} is to empower users to utilize system-level tools easily through proactive query recommendations. To fulfill this goal, \toolkit{} is carefully constructed to align with users' daily usage patterns. Specifically, as illustrated in Figure~\ref{fig:toolkit}, \toolkit{} includes 708 tools, comprehensively encompassing diverse functional domains like \textit{Display}, \textit{General System}, \textit{Media}, \textit{Communications}, \textit{Networks}, \textit{Traffic}, \textit{Health}, and others. By integrating these diverse categories, it comprehensively covers common system-level tool invocation demands in on-device scenarios—ranging from everyday tasks like "play music" to device maintenance actions such as "free up space". This well-structured preparation ensures that \modelname{} have access to a rich set of actionable tools, thus establishing a solid foundation for the subsequent recommendation of tool-invoking queries.

\textbf{Tool Retrieval.} 
Given the extensive scale of the \toolkit{}, it is impractical to include all available tools within the LLM's prompt during query recommendation. To address this limitation, we devise a context-aware tool retrieval mechanism, ensuring the model is conditioned on a subset of dialogue-relevant tools prior to making recommendations. As illustrated in Figure~\ref{fig:framework}(b), we first prepare textual descriptions for each tool and encode them into a vector database using the Qwen-3-embedding model~\cite{qwen3_emb}. During the inference phase, the user's dialogue history is similarly embedded, and a Qwen-3-reranker model is subsequently employed to filter and extract the top-$N$ most relevant tools according to the encoded descriptions. By employing this retrieval mechanism, \modelname{} is provisioned with a focused set of contextually relevant tools. The benefits are two-fold: 1) this design effectively mitigates the burden on the LLM's context window through reducing the number of tools provided to the LLM before recommendation. 2) Since only relevant tools are provided as input in \modelname{}, the generated tool-related queries naturally preserve a high relevance to the user's conversation history.


\subsection{Dual-level Preference Calibration}
\label{sec:dualc}
As depicted in Figure~\ref{fig:framework}(b), upon the model deployment, we gather online user logs and derive initial preference data from user click behavior. To further refine the quality of the preference data, we introduce a preference calibration mechanism from both the user and system sides.


\textbf{User-side Calibration.}
As formulated in Eq.(\ref{eq:kto_loss}), the standard KTO objective assumes equal importance for all samples by assigning uniform weights, which neglects the inherent behavioral biases across users with varying activity levels. In practice, the confidence of an interaction signal is relative to a user's natural click propensity. Intuitively, for highly active users who frequently interact, their deliberate non-clicks (negative samples) serve as a stronger indicator of irrelevance~\cite{DBLP:conf/cikm/XuanSSC25}. Conversely, for generally inactive or picky users, a rare click (positive sample) may heavily signify exceptional relevance and precise intent matching. To mitigate these behavioral disparities, we empirically propose a weighting strategy based on the user's click-through rate ($uctr$), which aims to heighten the model's sensitivity to the nuanced feedback of exceptionally active or inactive users, thereby properly calibrating the learning process. The dynamic weight $w_u$ for each sample is computed as follows:
\begin{equation}
    w_u = 
    \begin{cases}
    1 - \alpha \cdot \text{tanh}\left(\frac{uctr - \mu}{s}\right) & \text{if} \ Q_r\sim Q_r^+ \\
    1 + \alpha \cdot \text{tanh}\left(\frac{uctr - \mu}{s}\right) & \text{if} \ Q_r\sim Q_r^-
    \end{cases}
    \label{eq:user}
\end{equation}
where $\alpha$ is a hyperparameter to define the weight bounds, and $s$ is the standard error of all users' $uctr$. Given the long-tail distribution of $uctr$ in the real-world~\cite{user_fairness}, $\mu$ is empirically set to 0.07, the upper quartile of the $uctr$ distribution. Specifically, Eq.(\ref{eq:user}) is designed to dynamically scale the weights based on user activity: for active users ($uctr > \mu$), it amplifies the weight of negative samples while discounting positive ones; conversely, for inactive users ($uctr < \mu$), it naturally up-weights their rare positive clicks.


\textbf{System-side Calibration.} 
In the context of on-device intelligent assistants, a core user expectation is the rapid and accurate invocation of system-level utilities, such as freeing up storage or playing music. To prioritize this execution-oriented demand over casual chat, we introduce a system-side calibration to refine the raw preference data. Unlike standard alignment objectives that treat all logged queries equally, our calibration explicitly steers \modelname{} toward generating actionable queries $Q_r^t$---those capable of successfully triggering tools within the predefined \toolkit{}. To further align with real-world user needs, this mechanism assigns higher importance to queries associated with frequently utilized tools. Therefore, for a given recommended query $Q_r$, the system-side weight $w_s$ is formulated as follows:
\begin{equation}
    w_s = 
    \begin{cases}
        (1+\gamma)^{p^k} & \text{if} \ Q_r \sim Q_r^+ \ \& \ Q_r \sim Q_r^{t}
        \\
        0 & \text{if} \ Q_r \sim Q_r^- \ \& \ Q_r \sim Q_r^{t}
        \\ 
        1 & \text{else}
    \end{cases}
    \label{eq:system}
\end{equation}
where the hyperparameter $\gamma$ controls the maximum weight bound, and $p \in [0, 1]$ represents the normalized frequency percentile of the invoked tool. To modulate the sensitivity of the weight to click frequency, we introduce the hyperparameter $k$, which is empirically set to 3 based on a statistical analysis of our online interaction logs. Additionally, to consistently encourage the generation of executable tool-invoking queries, the second condition in Eq.(\ref{eq:system}) explicitly assigns a weight of $0$ to unclicked tool-invoking queries. This effectively masks these negative samples during training, ensuring that the model is not penalized for imperfect tool recommendations. Consequently, this design prevents the model from adopting a conservative strategy of favoring safe, general responses, thereby sustaining its proactiveness in invoking tools.

\subsection{Calibrated Preference Alignment}
In real-world online deployments, acquiring paired preference data—where both a chosen and a rejected response are annotated for the exact same input—is highly impractical. Consequently, our alignment stage relies on the KTO~\cite{kto} framework, which naturally accommodates unpaired feedback. Building upon the dual-level calibrated weights derived in Section~\ref{sec:dualc}, we propose a sample-level weighted KTO objective. The loss function is formulated as:
\begin{equation}
    \mathcal{L}= \mathbb{E}_{x,y \sim D} [w(Q_r)(1 - v(x, y; \beta))]
    \label{eq:final_loss}
\end{equation}
where the final sample weight $w(Q_r)$ is assigned as follows:
\begin{equation}
    w(Q_r)= 
    \begin{cases}
        \text{max}(w_u,w_s) \quad \text{if} \ Q_r \sim Q_r^+ 
        \\
        \text{min}(w_u,w_s) \ \quad \text{if} \ Q_r \sim Q_r^- 
    \end{cases}
\end{equation}

Crucially, this aggregation strategy is designed to balance optimization aggressiveness. For positive samples ($Q_r^+$), applying the $\max$ operator acts as an aggressive reward mechanism, ensuring that queries with either high user confidence or high system utility are strongly encouraged. Conversely, for negative samples ($Q_r^-$), the $\min$ operator serves as a conservative penalty, preventing the model from over-penalizing non-clicks unless both the user signal and system utility confidently indicate a negative preference.

\section{Experiment}
\subsection{Experimental Setup}
\textbf{Dataset.} We evaluate \modelname{} and other baselines through a large-scale online A/B test on OPPO Xiaobu, a leading mobile assistant platform with over 150 million monthly active users. To ensure a fair comparison, we allocate an equal volume of traffic from the main market to both the control and treatment groups. Prior to the formal experiment, both groups will be monitored for over 12 hours to verify that the distribution of user requests was consistent, ensuring that no initial bias existed between the two groups.

\textbf{Evaluation Metrics.} We employ three primary evaluation metrics: total clicks, Click-Through Rate (CTR), and relevance. Specifically, the relevance metric measures the proportion of recommended queries that are judged to be contextually aligned with the user's input. Given the substantial volume of daily user interactions, computing this metric across the entire online logs is computationally prohibitive. To estimate the relevance of each model, we randomly sample 1,000 recommendation instances per model from the online logs, which are subsequently evaluated for contextual relevance by Doubao-Seed-1.8~\footnote{\url{https://research.doubao.com/en/seed1_8}}.

\textbf{Reproducibility.} All experiments are conducted on a server equipped with 8 NVIDIA A100 GPUs and 200GB of RAM. We employ Qwen-3-14B~\cite{qwen3} as the base model and adopt \textit{Low-Rank Adaptation} (LoRA)~\cite{hu2022lora} for parameter-efficient fine-tuning. Specifically, the LoRA adapters are applied to all layers within the network with the rank set to 8. During the training phase, the batch size is set to 32. To ensure training stability, we utilize the AdamW optimizer with a learning rate of $5 \times 10^{-6}$, coupled with a cosine learning rate scheduler and a warmup ratio of $0.1$. Regarding the optimization objective, the hyperparameter $\beta$ in the KTO loss is set to 0.01, while the hyperparameters $\alpha$ and $\gamma$, which control the calibration weights $w_u$ and $w_s$, are set to $0.25$ and $1.25$, respectively.

\subsection{Main Online Results}
\label{sec:main}

\begin{table}[t]
\centering
\caption{Comparisons among different strategies. For online A/B test, each model is assigned with 5\% user traffic from the main market.}
\vspace{-0.1cm}
\label{tab:main}
\renewcommand{\arraystretch}{1.2}
\begin{small}
\begin{sc}
\begin{tabular}{lccc}
\toprule[1.1pt]
\textbf{Strategies} & \textbf{Click Number} & \textbf{CTR} & \textbf{Relevance} \\ \midrule[0.5pt]
\textbf{Base} & 1,063,499 & 0.3095 & \textbf{0.9710} \\
\textbf{SFT} & 1,069,529 & 0.3098 & 0.9590 \\
\textbf{Vanilla KTO} & 1,100,807 & 0.3167 & 0.9560 \\ \midrule[0.5pt]
\textbf{\modelname{}} & \textbf{1,113,871} & \textbf{0.3198} & 0.9570 \\
\text{Improve.} & +4.74\% & +3.32\% & -1.44\%\\
\bottomrule[1.1pt]
\end{tabular}
\vspace{-0.2cm}
\end{sc}
\end{small}
\end{table}

In this section, we compare \modelname{} with several other established alignment algorithms, including SFT and KTO~\cite{kto}. Note that, due to the specific characteristics of the online data, the acquisition of positive and negative sample pairs with identical inputs is impractical; consequently, algorithms such as DPO~\cite{dpo} and SimPO~\cite{simpo} are excluded from this comparison. We conduct an online A/B test from April 21 to April 27, 2026, allocating 5\% of the traffic to each model. Experimental results in Table~\ref{tab:main} reveal the following findings:

\begin{itemize}[leftmargin=*]
    \item Although both SFT and KTO show consistent gains over the base model, \modelname{} achieves the highest improvements in click number and CTR among all alignment methods evaluated, with increases of 4.74\% and 3.32\%, respectively. Notably, even when deployed to a mere 5\% of the total user traffic, this relative lift translates into a massive absolute gain of tens of thousands of additional user clicks.
    \item Compared to the base model, all alignment methods exhibit a decrease in relevance. We posit that prior to alignment, the base model strictly adhered to the historical context when generating queries. In contrast, aligned models are optimized to generate query recommendations that align with user preferences, rather than strictly pursuing content relevance.
    \item In particular, \modelname{} does not significantly compromise the relevance of its recommended queries to achieve these CTR improvements. Its relevance metric remained comparable to those of SFT and KTO, attaining a relatively high relevance score of 0.956. We attribute this preserved relevance to the proposed context-aware tool retrieval mechanism, which effectively ensures that the generated queries are still strictly grounded in the user's current dialogue context.
\end{itemize}

\subsection{Offline Comparisons}

\begin{figure}[t]
    \centering
    \includegraphics[width=3.1
    in]{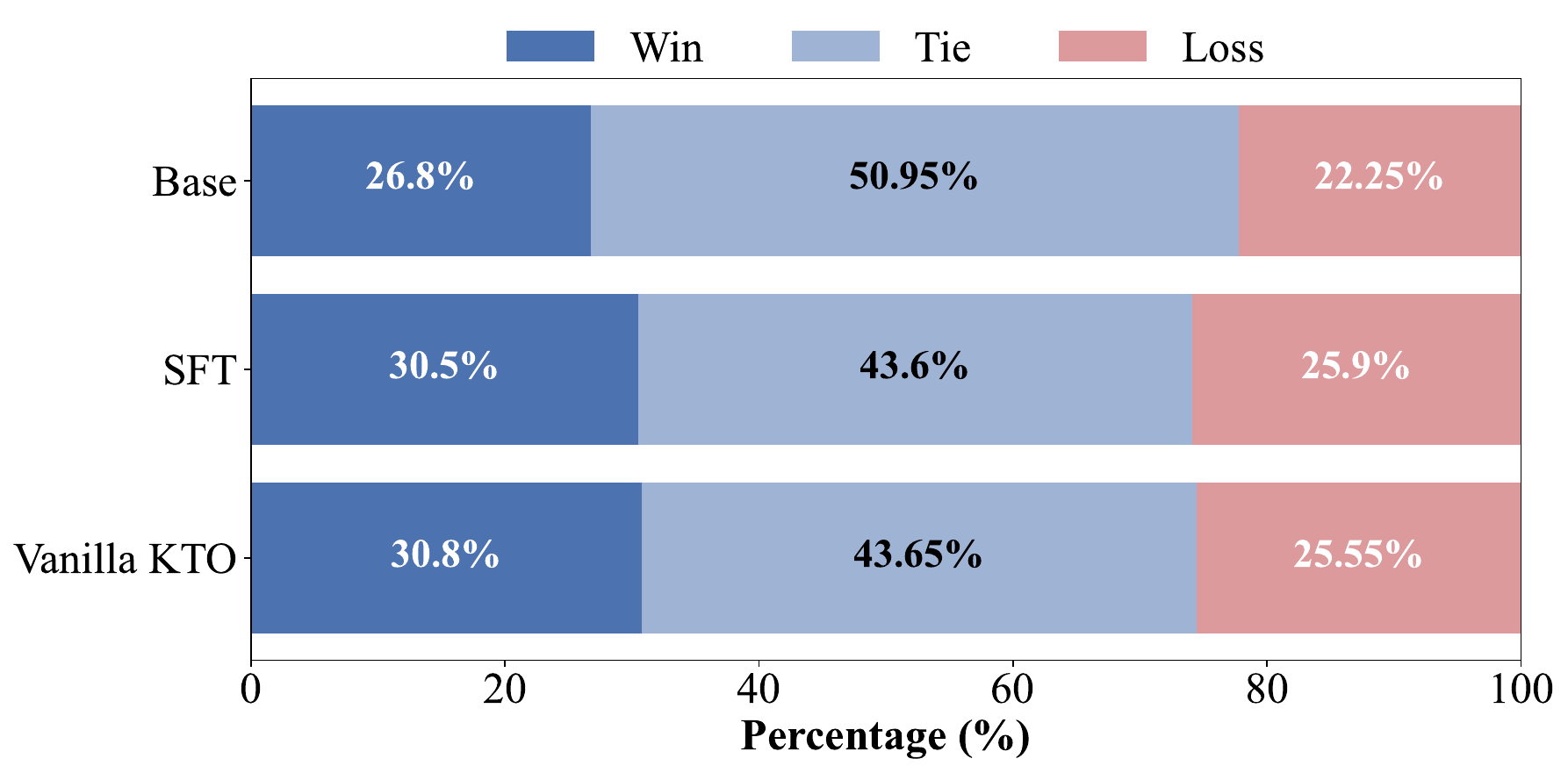}
    \caption{Offline comparison among \modelname{} and baselines}
    \vspace{-0.3cm}
    \label{fig:gsb}
\end{figure}

To further validate the effectiveness of \modelname{} beyond online metrics, we conduct a comprehensive offline analysis. Specifically, we randomly sample 2,000 real-world user dialogue histories from online logs to serve as the input contexts of \modelname{} and baselines. To assess the quality of the generated query recommendations, we adopt an LLM-as-a-judge paradigm~\cite{judgelm, paperbench}, utilizing Doubao-Seed-1.8 to perform strict pairwise evaluation. For each given context, the evaluator determines whether \modelname{} achieves a Win, Tie, or Loss against the baseline models based on the helpfulness of tool invocation and the diversity of the recommended queries.

As illustrated in Figure~\ref{fig:gsb}, \modelname{} consistently outperforms all three baselines, achieving higher win rates than loss rates. 
Interestingly, we observe a notably high proportion of "Tie" outcomes across all three comparison pairs, ranging from 43.6\% to roughly 50\%. We attribute this phenomenon to the models' comparable capabilities in ordinary contextual conversations. In cases where users have no tool-invoking demands, the query generation heavily relies on general conversational abilities. Given that all models in Figure~\ref{fig:gsb} share the same foundational model (Qwen-3-14B), they are all highly capable of generating the exact correct queries for these routine tasks, leaving minimal room for improvement.

\subsection{Ablation Study}

\begin{table}[t]
\centering
\caption{Ablation study. Each model is assigned with 2\% user traffic from the main market.}
\vspace{-0.1cm}
\label{tab:ablation}
\renewcommand{\arraystretch}{1.05}
\begin{small}
\begin{sc}
\resizebox{\linewidth}{!}{
    \begin{tabular}{c | c c c | c c}
        \toprule
        \multicolumn{4}{c|}{\textbf{Model Variants}} & \multirow{2}{*}{\textbf{Click Number}} & \multirow{2}{*}{\textbf{CTR}}\\
        \cmidrule{1-4}
        ID & $w_u$ & static $w_s$ & Dynamic $w_s$ & \\
        \midrule
        1       & $\times$   & $\times$   & $\times$   & 423,561 & 0.3051 \\
        2       & \checkmark & $\times$   & $\times$  & 434,084 & 0.3090 \\
        3       & $\times$ & \checkmark & $\times$  & 429,931 & 0.3080 \\
        4       & \checkmark & \checkmark & $\times$  & 446,527  & 0.3162 \\ \midrule
        \modelname{} & \checkmark & $\times$ & \checkmark & \textbf{458,334} & \textbf{0.3226} \\
        \bottomrule
    \end{tabular}}
\end{sc}
\end{small}
\end{table}

In this section, we design several model variants to evaluate the effectiveness of the components within \modelname{}. Specifically, we examine the following four variants: (1) vanilla KTO (i.e., with no preference calibration applied); (2) user-side calibration only; (3) system-side static weight calibration only (i.e., calibration weights are independent of tools' frequency); and (4) a combination of both user-side and system-side static calibration. We conduct an online A/B test over 7 days (April 28 to May 4, 2026), allocating 2\% of the traffic to each model. The experimental results are presented in Table~\ref{tab:ablation}, where we can have the following observations:

\begin{itemize}[leftmargin=*]
    \item By comparing Variant 1 with Variants 2 and 3, we observe that implementing either user-side calibration or system-side static calibration independently leads to notable improvements in both the CTR and the total number of clicks. This explicitly demonstrates the individual effectiveness of both calibration strategies in refining preference signals.
    \item The comparisons between Variant 4 and its predecessors (Variants 2, 3) reveal that user-side and system-side calibrations do not conflict with one another. On the contrary, the two mechanisms are highly complementary, and their joint application yields substantial performance gains over using either in isolation.
    \item By comparing Variant 4 with the complete \modelname{},  we find that upgrading from static to frequency-based system-side calibration yields a substantial performance leap. We posit that calibrating weights based on tool frequency encourages the model to prioritize tool-related queries that receive frequent user interaction, thereby resulting in superior overall performance.
\end{itemize}

\subsection{Hyperparameter Analysis}

\begin{figure}[t]
    \centering
    \includegraphics[width=3.15
    in]{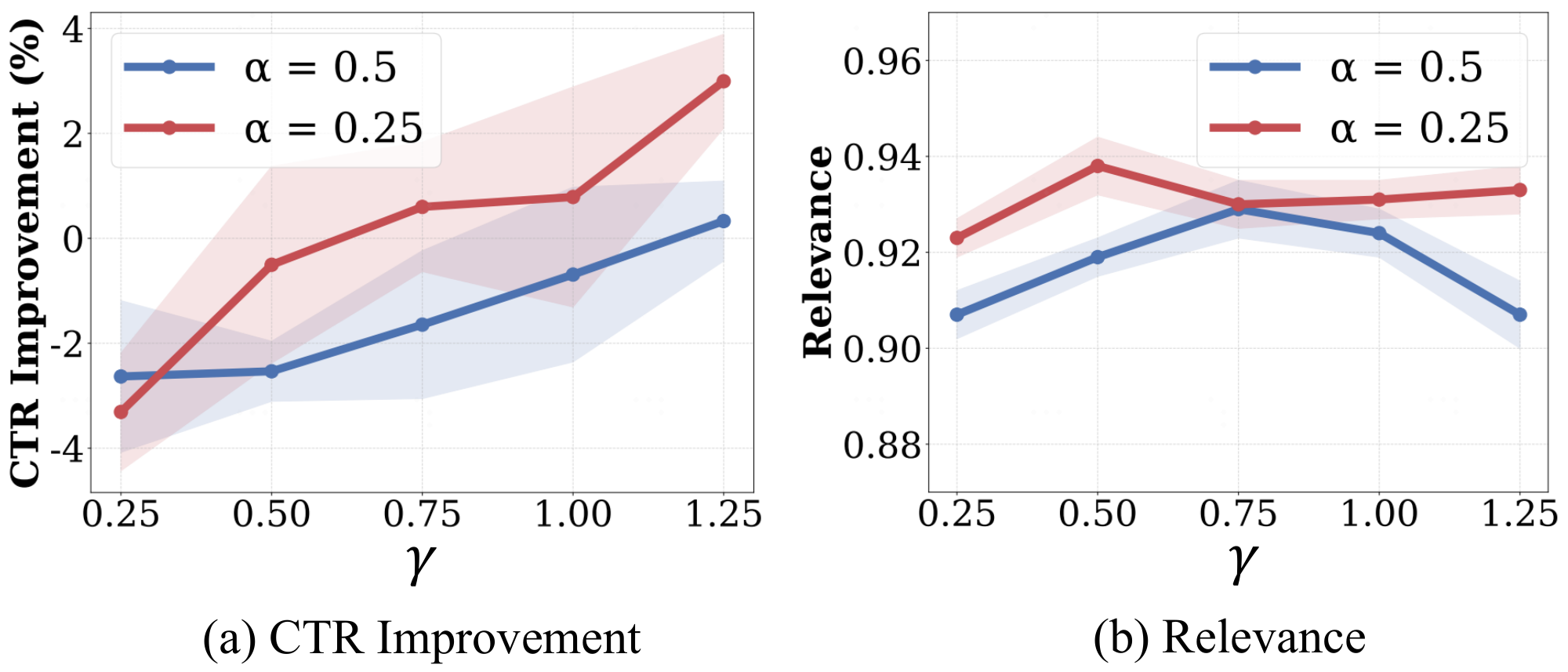}
    \vspace{-0.2cm}
    \caption{The impact of $\alpha$ and $\gamma$ on \modelname{}}
    \label{fig:hyper}
\end{figure}

In this subsection, we investigate the impact of two critical weight hyperparameters, $\alpha$ and $\gamma$, within \modelname{}. Both weights control the range of preference calibration. Specifically, due to the limited availability of concurrent online experimental slots, we conducted online A/B tests by allocating 2\% of live traffic across two distinct periods: April 4-6 and April 11-13, 2026. To mitigate inherent temporal fluctuations in CTR across these different periods, we report the relative CTR improvements of various hyperparameter configurations against the vanilla KTO baseline. As illustrated in Figure~\ref{fig:hyper}(a), the relative CTR improvement exhibits a steady upward trend as $\gamma$ increases. Conversely, relevance in Figure~\ref{fig:hyper}(b) follows a different trajectory: it initially rises, but subsequently plateaus or even declines. This observation matches our expectation, since excessive emphasis on generating tool-related queries inevitably undermines contextual coherence and relevance. Empirical results indicate that the configuration [$\alpha=0.25, \gamma=1.25$] strikes an optimal balance, yielding highly competitive performance.

\subsection{Granular Performance Evaluation}

To further validate the effectiveness of \modelname{}, in this section we aim to present a granular evaluation of its performance across following perspectives.

\begin{figure}[t]
    \centering
    \includegraphics[width=3.15
    in]{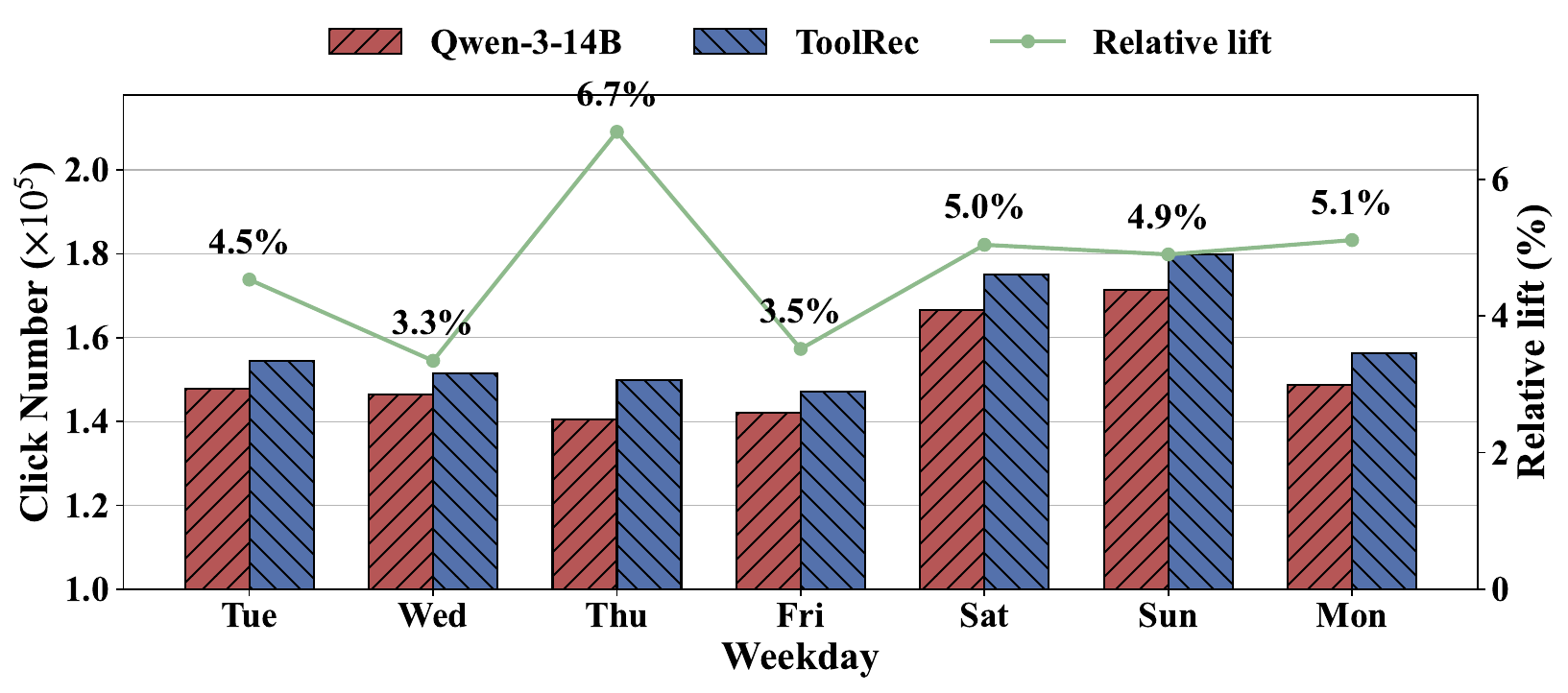}
    \vspace{-0.1cm}
    \caption{The daily click number gains of \modelname{}}
    \vspace{-0.1cm}
    \label{fig:daily}
\end{figure}

\textbf{Day-wise Performance Analysis.} We first examine the temporal stability of \modelname{}. Figure~\ref{fig:daily} illustrates the distribution of daily click numbers alongside the relative performance lift of \modelname{} against the base model over a one-week period (April 21 to April 27, 2026). Consistent with the experimental setup in Section~\ref{sec:main}, 5\% of the total traffic was allocated to each model. As shown in the results, interaction volume is significantly higher on weekends than on weekdays, likely due to users spending more leisure time on their devices, thereby increasing their engagement with the assistant. Despite these daily traffic fluctuations, \modelname{} consistently outperforms the base model, achieving a stable relative lift in click numbers ranging from 3.3\% to 6.7\%. This demonstrates that the effectiveness of our proposed method is robust against temporal variations.

\textbf{Stratified User Analysis.} To evaluate the effectiveness of \modelname{} across varying levels of user engagement, we conduct a stratified analysis by partitioning users into high-activity and low-activity cohorts based on their historical $uctr$. As detailed in Table~\ref{tab:high_low_users}, we compare the performance of all models across these two distinct user segments. The results demonstrate that \modelname{} consistently achieves highly competitive results, ranking within the top two for both click numbers and CTRs across both user groups. We attribute this robust performance to the proposed user-side preference calibration mechanism. By dynamically adjusting sample weights according to individual click propensities, this approach effectively mitigates inherent behavioral noise and yields higher-quality preference data.

\begin{table}[t]
    \centering
    \renewcommand{\arraystretch}{1.1}
    \caption{Performance comparison across high-ctr and low-ctr user groups. Each model is assigned with 5\% user traffic.}
    \label{tab:high_low_users}
    \scshape 
    \resizebox{\linewidth}{!}{
    \begin{tabular}{l cccc}
        \toprule
        \multirow{2}{*}{{\textbf{Models}}} & \multicolumn{2}{c}{{\textbf{Click Number}}} & \multicolumn{2}{c}{{\textbf{CTR}}} \\
        \cmidrule(lr){2-3} \cmidrule(lr){4-5}
        & \makecell{High-ctr \\ Users} & \makecell{Low-ctr \\ Users} & \makecell{High-ctr \\ Users} & \makecell{Low-ctr \\ Users} \\
        \midrule
        {Base} & 467,449 & 27,372 & 0.8330 & \underline{0.0939} \\
        {SFT} & 470,639 & \textbf{27,568} & 0.8320 & 0.0938 \\
        {Vanilla KTO} & \underline{493,373} & 26,584 & \textbf{0.8417} & 0.0904 \\ \midrule
        {\modelname{}} & \textbf{509,043} & \underline{27,519} & \underline{0.8358} & \textbf{0.0945} \\
        \bottomrule
    \end{tabular}}
\end{table}

\textbf{Query Type Distribution Analysis.}
A fundamental objective of \modelname{} is to incentivize the generation of actionable, tool-invoking queries over standard general queries. To evaluate this capability, we statistics the distribution of recommended queries using online serving logs collected from April 21 to April 27, 2026, where each model was deployed to a $5\%$ slice of the total user traffic. Specifically, we report the relative shift in the proportion of tool-invoking queries generated by various models compared to the base model. As illustrated in Table~\ref{tab:query_type}, \modelname{} achieves the most substantial relative uplift of $1.44\%$, explicitly validating the effectiveness of our proposed system-side preference calibration mechanism. Note that, considering the massive volume of daily user interactions in OPPO Xiaobu, a $1.44\%$ proportional shift is still significant. Even under the restricted $5\%$ traffic allocation, this relative uplift can transform into an impact on around 20,000 user requests.

\begin{table}[t]
\centering
\caption{Relative improvement in the proportion of tool-invoking queries during the online A/B test. Each model is assigned 5\% of the user traffic from the main market.}
\label{tab:query_type}
\renewcommand{\arraystretch}{1.2}
\begin{small}
\begin{sc}
\begin{tabular}{lc}
\toprule[1.1pt]
\textbf{Strategies} & \textbf{Relative Improvement} \\ \midrule[0.5pt]
\textbf{Base} & --  \\
\textbf{SFT} & -0.19\%  \\
\textbf{Vanilla KTO} & +0.45\% \\ \midrule[0.5pt]
\textbf{\modelname{}} & \textbf{+1.44\%} \\
\bottomrule[1.1pt]
\end{tabular}
\end{sc}
\end{small}
\end{table}


\subsection{Case Study}

\begin{figure}[t]
    \centering
    \includegraphics[width=3.3
    in]{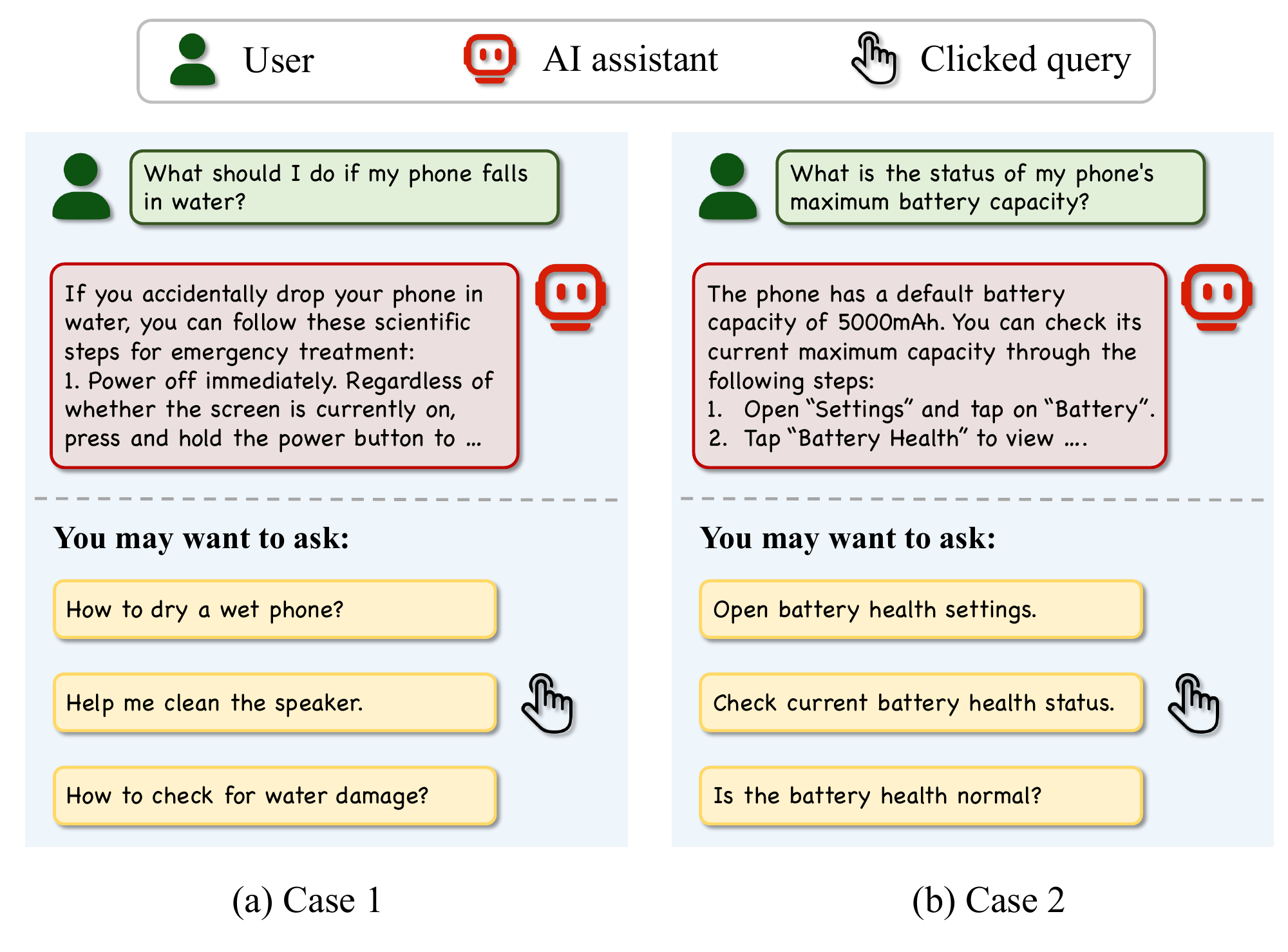}
    \vspace{-0.15cm}
    \caption{Two cases where \modelname{} successfully provides tool-invoking queries that users click. The content is translated into English for better understanding.}
    \vspace{-0.2cm}
    \label{fig:case}
\end{figure}

To qualitatively illustrate the practical advantages of \modelname{} in real-world scenarios, we present two representative cases in Figure~\ref{fig:case}, where the model successfully anticipates user needs and recommends highly relevant tool-invoking queries.

In the first case, a user seems to seek emergency assistance after accidentally dropping the phone in water. Alongside standard text-based troubleshooting instructions, \modelname{} proactively recommends a set of follow-up queries, notably including a tool-invoking query: \textit{"Help me clean the speaker."}. In such a time-sensitive context, directly triggering the device's built-in water ejection tool is substantially more practical and urgent than suggesting generic informational queries. The user's subsequent click on this specific recommendation also supports this claim.

The second case involves a user inquiring about their device's maximum battery capacity, which typically implies an underlying concern regarding battery degradation or rapid power drain. After providing a step-by-step textual guide on how to navigate the phone's interface, \modelname{} immediately surfaces tool-invoking queries such as \textit{"Open battery health settings"} and \textit{"Check current battery health status."}. The user's interaction with the latter confirms that when assessing device status, users highly favor one-click executable actions over manual navigation.

Collectively, these case studies demonstrate that \modelname{} effectively bridges the gap between conversational understanding and system-level execution. By accurately grounding the dialogue context in the device's capability space, the model consistently provides actionable recommendations that align with genuine user intents.

\section{Conclusion}

In this paper, we present \modelname{}, a query recommendation framework engineered specifically for on-device intelligent assistants. Unlike conventional chatbots, we empirically find that in on-device intelligent assistants, users exhibit a stronger preference for actionable, system-level operations. To facilitate this, we first build \toolkit{}, a comprehensive repository of 708 system tools, coupled with a context-aware tool retrieval mechanism that guarantees the relevance of the recommended queries. To accurately capture authentic user intent, we further implement a dual-level preference calibration mechanism that filters out behavioral noise on the user side while explicitly amplifying the importance of tool-invoking actions on the system side. By optimizing the model via a sample-level weighted KTO approach using this refined feedback, \modelname{} seamlessly aligns with genuine user demands. Online deployment on OPPO Xiaobu with over 150 million monthly active users verifies the effectiveness of \modelname{}, showing substantial gains in both CTR and click volume against competitive baselines.



\bibliographystyle{ACM-Reference-Format}
\bibliography{sample-base}

\newpage
\appendix

\end{document}